\documentstyle[aps]{revtex}
\def\equa{\mathrel{\raise-7pt\hbox to 26pt{\raise
11pt\hbox{\mbox{min}}\hss{\small{$j = 1,2$}}}}V(|\vec{r}_{i}-\vec{R}_{j}|)}
\begin{document}

\baselineskip=7mm

\title{\bf Systematic Theoretical Search for Dibaryons in a Relativistic  
Model}

\vspace{0.1in}

\author{T. Goldman}
\address{Theoretical Division, Los Alamos National Laboratory, Los  
Alamos,
NM 87545 USA}

\author{K. Maltman}
\address{Department of Mathematics and statistics, York University,
 North York, Ontario, Canada M3J 1P3}

\author{G. J. Stephenson Jr}
\address{Department of Physics, University of New Mexico, Albuquerque,
NM 87131 USA }

\author{Jia-Lun Ping}
\address{Department of Physics , Nanjing Normal University, Nanjing,
210097, China}

\author{Fan Wang}
\address{Department of Physics and Center for Theoretical Physics,
Nanjing University, Nanjing, 210008, China }

\maketitle

\vspace{0.2in}

\begin{abstract}
A relativistic quark potential model is used to do a systematic search
for quasi-stable dibaryon states in the $u$, $d$, and $s$ three flavor
world. Flavor symmetry breaking and channel coupling effects are
included and an adiabatic method and fractional parentage expansion
technique are used in the calculations. The relativistic model predicts
dibaryon candidates completely consistent with the nonrelativistic model.
\end{abstract}

\newpage

\section{Introduction}

Since Jaffe predicted the first dibaryon, the H particle\cite{1}, there
have been many experimental and theoretical efforts to search for
dibaryons. Up to now, there are no experimentally well established
dibaryon states except the deuteron. Theoretically almost all QCD
models, including lattice QCD, predict that there should be dibaryon
states\cite{2}.  However no model has yet achieved an acceptable level of
quantitative reliability. Either the model Hamiltonian is
oversimplified or the model Hilbert space is rather restricted, or
both. We developed a nonrelativistic model, which we termed the quark
delocalization color screening model(QDCSM)\cite{3,4,5}. The model
Hamiltonian and Hilbert space are reasonable enough to yield a
qualitatively correct N-N interaction including both repulsive core and
intermediate range attraction and to fit all of the four ($IJ$)=(01)
(10) (00) and (11) channel phase shifts qualitatively. This model
reproduces an almost correct deuteron state and verifies dynamically
that there are two kinds of dibaryons\cite{6}. One kind is a loosely bound,
low spin ($J$), two baryon state with only slight quark delocalization,
of which the deuteron is prototypical. Another kind is a tightly bound,
high spin, six quark state with significant quark delocalization; here,
the $d^{\star}$($IJ$=03)\cite{6} is a prime example. (We do not discuss, 
however, potential dibaryons with more involved internal structure, 
such as those with non-valence internal structure, such as would 
correspond to NN$\pi$ bound states\cite{garcz}, or with orbital 
excitations of the valence quarks\cite{faess}.) 

It is believed that systems of light ($u, d$ and $s$) quarks require
relativistic dynamics. Therefore it is better to have a relativistic
model calculation to check if those dibaryon states predicted by the
nonrelativistic model(QDCSM) are robust against relativistic effects. A
relativistic quark potential model(LAMP)\cite{7} has been developed for
this purpose and the fractional parentage technique developed for the
nonrelativistic dibaryon calculation has been extended to the
relativistic case as well\cite{8}.  Together, these make a systematic
relativistic quark potential model dibaryon search feasible and the
results are reported here. 

\section{The relativistic quark potential model}

We neglect the small current quark mass for $u$ and $d$ and treat them
as massless, $m_{u}=m_{d}=m=0$. For a single baryon, we assume a
relativistic Hamiltonian with a scalar linear confinement\cite{7},
\begin{equation}
H(B)=\sum^{3}_{i=1} H_{i}+\sum_{i<j} H_{ij}
\end{equation}
\begin{equation}
H_{i}=\vec{\alpha}_{i}\cdot\vec{p}_{i}+\beta_{i}(m_{i}+V(r_i))
\end{equation}
\begin{equation}
V(r_i)=k^2 (r_i-r_0)
\end{equation}
\begin{equation}
H_{ij}=g_s(m_i)g_s(m_j)\frac{\vec{\lambda_i}\cdot\vec{\lambda_j}}{4}
\vec{\alpha}_i\cdot\vec{\alpha}_j A e^{-\nu(\vec{r_i}-\vec{r_j})^2}
\end{equation}
where $\alpha_i, \beta_{i}$ are Dirac matrices, $\vec{\lambda}_i$ is
the color SU(3) generator, $m_i$ is the quark mass. $V(r)$ is a
phenomenological confinement potential, which we assume is a Lorentz
scalar and $r_i$ is the modulus of the three-coordinate of the $i$-th
quark relative to the origin. The color Coulomb interaction due to
gluon exchange is assumed to be absorbed in the form of Eq.(3) even
though it is a Lorentz vector. This is done to simplify the numerical
calculation. The parameters $k^2$=0.9~GeV/fm and $r_0=0.57$~fm are
fixed by the condition that the eigenenergy $E$ of the following Dirac
equation\cite{9}.
\begin{equation}
H_{i}\psi_{\sigma_i}(\vec{r}_i)=E\psi_{\sigma_i}(\vec{r}_i)
\end{equation}
is one third of the average of the nucleon (N) and Delta($\Delta$)
masses, $E=\frac{1}{6}(N+\Delta)$, for massless quarks. The lowest
energy wavefunction solution is:

\[\psi_{\sigma_i}(\vec{r_i})=\left(\begin{array}{c}
                    \phi_{u}(r_i) \\
                    -i\vec{\sigma_i}\cdot\hat{r}_i\phi_{l}(r_i)
                    \end{array}  \right) \]
\[
\phi_{u}(r)=P_{u}(kr)e^{-k^2 r^2/2}
\]
\begin{equation}
\phi_{l}(r)=P_{l}(kr)e^{-k^2 r^2/2}
\end{equation}
\[
P_{u}(kr)=1+0.04287 kr+0.00457 (kr)^2+\cdots
\]
\[
P_{l}(kr)=-kr(0.46330-0.08767 kr+\cdots)
\]
$H_{ij}$ is the color magnetic interaction due to gluon exchange,
assumed to keep the form of single gluon exchange. A Gaussian gluon
propagator is adopted for the color magnetic interaction to simulate
the confinement property of a color gluon on the one hand and to
simplify the numerical calculation on the other hand.
$\nu=(\frac{0.2GeV}{\hbar c})^2=1.0$ fm$^{-2}$ is chosen to be about
the size of a hadron. The combination $g^{2}_{s}(m=0) A=-1.350$~GeV is
fixed by $N-\Delta$ mass difference, where $A$ is the matrix element of
the spatial part of the (color) current-current interaction matrix
element. Note that the quark-gluon effective coupling is assumed to
vary with a scale related to the quark mass.

Note that, having so-modified the form of single gluon exchange, this 
two-body-interaction Hamiltonian could equally well be taken to represent 
the effects of instantons\cite{shuryak}. The color structure, limited 
range, and quark-mass-dependent effective strength are all also 
features found in the effective quark-quark interactions produced 
by the propagation of quarks in the presence of instantons. The 
diquark correlations expected in instanton models are thus implicitly 
implemented by this $H_{ij}$.

The single baryon ground state wave function (WF) is assumed to be a
product of single quark WFs,
\begin{equation}
\psi_{B}(123)=\chi_{c}(123)\left[ \prod^{3}_{i=1}\psi_{\sigma_{i}}(\vec{r}_i)
\eta_{f_i}(i) \right]_{SIJ}
\end{equation}
where $\chi_{c}$(123) is a three quark color singlet state.
$\eta_{f_i}$ is the single quark flavor WF. [~]$_{SIJ}$ means the
individual quark spin-flavors are coupled to total baryon strangness
$S$, isospin $I$ and spin $J$.

The strange quark mass $m_{s}=307$~MeV is determined by the $\Lambda-N$
mass difference. (Although a factor of about two larger than
conventional in quark models, this difference can be traced to a virial
theorem factor of one-half which arises from the use of a Dirac scalar
potential.)  The ratio, ${g_s(m_s)}/{g_s(0)}$, is determined by the
overall fit of the octet and decuplet baryon masses. The fitted baryon
masses and the value of $g_{s}(m_s)/g_s(0)$ are shown in Table 1.

\begin{center}
Table 1. Baryon masses. $g_{s}(m_s)/g_s(0)$=1/1.26

\begin{tabular}{||lcccccccc||} \hline
   & N     & $\Delta$ & $\Lambda$ & $\Sigma$ & $\Xi$ & $\Sigma^{\star}$  
&
$\Xi^{\star}$ & $\Omega$ \\
exp.   & 939 & 1232 & 1116 & 1193 & 1318  & 1385  & 1533 & 1675  \\
theor. & 939 & 1232 & 1116 & 1182 & 1333  & 1376  & 1527 & 1683  \\  
\hline
\end{tabular}
\end{center}

For a two baryon system, we change the Hamiltonian slightly from that
given in Eq.(1): $i$ now runs from 1 to 6 and the confinement potential
is replaced by
\begin{equation}
V_{2}(r_{i}) = \begin{array}{c}
{\mbox{{\large min}}} \\ 
{\mbox{{\footnotesize J=1,2}}} \\ 
\end{array}
\{ \, V( \, | \, \vec{r}_{i} - \vec{R}_{J} \, | \, ) \,\} 
\end{equation}
where $\vec{R}_{1}$ and $\vec{R}_{2}$ are two baryon `centers'\cite{7} and
$V(r_{i})$ is given in Eq.(3). This confinement potential is devised to
simulate the effect that, as the two baryons are brought close
together, the quark matter density in between them will increase
gradually, causing the QCD vacuum to change (also gradually) from
nonperturbative to perturbative. As one quark is removed to a distance
large compared to the separation of the two centers, the confining
potential should be the same as for an isolated baryon. These effects
are achieved by truncating the value of the confining potential on the
midplane between the two baryon centers, $\vec{R}_{1}$ and
$\vec{R}_{2}$.

The two baryon WF is assumed to be a linear combination of the
following Dirac cluster WFs,
\begin{equation}
\Psi_{\alpha K}(q^6)= {\cal A} [\psi(B_{1})\psi(B_{2})]_{SIJ}
\end{equation}
\begin{eqnarray}
\psi(B_{1}) & = & \chi_{c_1}(1 2 3) \left[ \prod^{3}_{i=1}
  \psi^{L}_{\sigma_i}(\vec{r}_i) \eta_{f_i}(i) \right ]_{S_1I_1J_1}
  \nonumber \\
\psi(B_{2}) & = & \chi_{c_2}(4 5 6) \left[ \prod^{6}_{j=4}
  \psi^{R}_{\sigma_j}(\vec{r}_j) \eta_{f_j}(j) \right]_{S_2I_2J_2}
\end{eqnarray}
\begin{eqnarray}
\psi^{L}_{\sigma_i}(\vec{r}_i) & = & (\psi_{\sigma_i}(\vec{r}_i-\vec{R}_1)
+\epsilon(x)\psi_{\sigma_i}(\vec{r}_i-\vec{R}_2))/N(x)  \nonumber \\
\psi^{R}_{\sigma_j}(\vec{r}_j) & = & (\psi_{\sigma_j}(\vec{r}_j-\vec{R}_2)
+\epsilon(x)\psi_{\sigma_j}(\vec{r}_j-\vec{R}_1))/N(x)
\end{eqnarray}
\begin{eqnarray}
x & = & |\vec{R}_1-\vec{R}_2| \nonumber \\
N^2(x) & = & 1+\epsilon^2(x)+2\epsilon(x) \left\langle
 \psi_{\sigma_i}(\vec{r}_i-\vec{R}_1)|\psi_{\sigma_i}(\vec{r}_i-\vec{R}_2)
 \right\rangle
\end{eqnarray}
$\cal A$ is the normalized antisymmetrization operator. $\alpha=(SIJ)$
and $K$ represent the other quantum numbers related to $B_{1}$ and
$B_{2}$, $\chi_{c_1}$, and $\chi_{c_2}$ are both the three quark color
singlet states.

A physical two baryon state with quantum number $\alpha=(SIJ)$ is
\begin{equation}
\Psi_{\alpha}(q^6)=\sum_{K} C_{\alpha K}\Psi_{\alpha K}(q^6)
\end{equation}
The summation $K$ runs over all the possible two baryon states
consisting of the octet and decuplet baryons listed in Table I. This is
sufficient as it spans the space of relevant asymptotic states.

The six quark Hamiltonian is first diagonalized in the Dirac cluster
basis space given by Eq.(13). To simplify the six quark Hamiltonian
matrix element calculations, a relativistic extension of the fractional
parentage expansion method is used\cite{8}. The one-body and two-body
matrix elements are calculated by a program developed in the
preliminary relativistic dibaryon calculations\cite{7,10}.

The lowest eigen-energy obtained in the diagonalization is $x$ and
$\epsilon(x)$ dependent, and $\epsilon(x)$ is varied at each $x$ to
obtain the minimum.  The difference between the minimum at $x$ and
infinite separation is assumed to be the effective potential of the two
baryons. (The minimum at $x=3$~fm is already negligibly different from
values at larger separation.) We expect that spurious effects such as
center-of-mass (c.m.) motion largely cancel out in the effective
potential due to the process described above. Therefore, we have
neither subtracted the c.m. motion energy separately for a single
baryon nor for the two baryons together. Following Ref.\cite{6}, we use the
following expression to calculate the dibaryon masses,
\begin{equation}
M_{\alpha}(q^6)=(M_1+M_2)_{\alpha}+ \mbox{min} [V_{\alpha}(x)] +  
\frac{3}{4
\mu_{\alpha}} \frac{\hbar^2}{x^2_0}
\end{equation}
where $\mbox{min} [V_{\alpha}(x)]$ is the minimum of the effective
interaction $V_{\alpha}(x)$, and $(M_1+M_2)_{\alpha}$ is the channel
weighted experimental two baryon mass,
\begin{equation}
(M_1 + M_2)_{\alpha} = \sum_{K} | C_{\alpha K}|^2 < \Psi_{\alpha K}
(q^6) | \Psi_{\alpha K} (q^6) > (M_1 + M_2)_{\alpha K},
\end{equation}
at the point $x_0$ at which the minimum occurs.

A zero point oscillation energy of $\frac{3 \hbar^2}{4 \mu_{\alpha}
x^2_0}$ has been taken into account as explained in Ref.\cite{6}.  Even
though the internal motion of the quarks is relativistic, we assume the
relative motion of the two baryon centers still can be approximated as
a nonrelativistic oscillation around the equilibrium separation $x_0$,
where $\mu_{\alpha}$ is the weighted reduced channel mass,
\begin{equation}
\mu_{\alpha} = \sum_{K} | C_{\alpha K} |^2 < \Psi_{\alpha K} (q^6) |
\Psi_{\alpha K} (q^b) > \left( \frac{M_1 M_2}{M_1 + M_2} \right)_{\alpha K} .
\end{equation}

This model has been used to calculate the binding energies of $^4$He
and $^3$He, for which the reasonable values 19~MeV ($^4$He),
3.8~MeV ($^3$He) have been obtained without any adjustable
parameters\cite{7}.

\section{Results}

A systematic dibaryon search has been done in the $u, d$ and $s$ three
flavor world. Both channel coupling and flavor symmetry breaking
effects have been taken into account. The interesting dibaryon states
are listed in Tables 2a,b.  Table 2a lists `Model A' results, where an
SU(3) flavor symmetric WF is used and the one body energy difference
due to $m_{s}\neq m$ have also been neglected.  The two body matrix
elements are assumed to obey the (empirically consistent) relations
\begin{eqnarray}
\langle sl|H_{ij}|sl\rangle  & = & \frac{2}{3} \langle ll|H_{ij}|ll\rangle
\nonumber \\
\langle ss|H_{ij}|ss\rangle  & = & \frac{4}{9} \langle ll|H_{ij}|ll\rangle
\end{eqnarray}
$\langle ll|H_{ij}|ll\rangle = A$ is calculated using the massless
quark WF (Eq.6) operator matrix element multiplied by the overall
constant such that $g^2_{s}(m=0)$A$ = -1.350$~GeV.  The notation sl(ss)
means strange-nonstrange (strange-strange) two body matrix elements.
Table 2b lists the `Model B' results where the one-body energy for the
$u, d$ and $s$ quarks and the two-body matrix elements are both
calculated with the WF obtained from Eq.(5) with $m$=0 and $m_{s} =
307$~MeV, respectively. The overall constants, $g^2_{s}(0)$A$ =
-1.350$~GeV and $g_{s}(0) g_{s}(m_{s})$A and $g^2(m_s)$A appear in
$\langle ll|H_{ij}|ll\rangle $, $\langle sl|H_{ij}|sl\rangle $ and
$\langle ss|H_{ij}|ss\rangle $ respectively to obtain the full flavor
dependence of the two body matrix elements. The value $g_{s}(m_s) =
g_{s}(0)/1.26$ determined by the ground state baryon masses has been
used.

In principle we should use an SU(3) flavor symmetry breaking WF to
study the flavor symmetry breaking effects. We use these two model
results to show the uncertainty of the flavor symmetry breaking effect
in our model. The largest difference is for the $\alpha = ( -600 )$
state where the model A and B results have 140~MeV difference.  Except
for this extreme strangeness case, the largest difference is about
60~MeV corresponding to $\alpha = (-3 \frac{1}{2} 2)$. The channel
coupling effects are similar to the nonrelativistic case.

The most prominent feature is that the relativistic model yields
dibaryon masses in the whole $u, d, s$ three flavor world quite similar
to those of the nonrelativistic model. This is shown in Table 3, where
NA, NB, RA and RB stand for nonrelativistic color screening version A
and B\cite{6} and the relativistic model A and B (with channel coupling and
flavor symmetry breaking, $ccb$) results, respectively. The similarity
is true also for states omitted from these tables.  This might be
considered surprising as the relativistic and nonrelativistic model are
not only different in kinematics but also in the details of the
confinement mechanism. We take this similarity of results as a mutual
confirmation of the stability to the model details of the estimated
dibaryon masses in these two models.

Comparing the relativistic~(R) and nonrelativistic~(N) model estimates
on the dibaryon masses more carefully, one finds there are differences
fluctuating from case to case.  

In the nonstrange sector, the R model masses are higher than the N
ones.  In the R model, the deuteron is unbound and the model $\alpha$ =
(003) state is $\sim$60~MeV higher than the N one.  For the strangeness
-1 sector, the N and R models give almost the same dibaryon masses.

Beginning at strangeness -1 and generally increasing with increasing
strangeness, the R model gives larger binding.  Except for the $\alpha$
= (-220) state, where the R model mass is 10~MeV higher than the N
model mass, in all the other cases, R masses are smaller than N masses,
ranging from -25~MeV for the $\alpha$ = (-400) state to -220~MeV for
the $\alpha$ = (-600) and $\alpha$ = (-3$\frac{1}{2}$2) states.  The
$\alpha$ = (-600) state is a special example, where the two versions of
the R model themselves have a 140~MeV difference, while all others
differ by less than $\pm$60~MeV.  However it is interesting to note
that skyrmion model practitioners have predicted the same strangeness
-6 dibaryon candidate\cite{11}.

This general trend of differences between the R and N models may be
understood in the following way: In the N model fit to the octet and
decuplet baryon masses, the theoretical octet masses are larger than
experiment while the decuplet masses are smaller. This indicates that
the color magnetic interaction decreases (theoretically) more than
warranted from the nonstrange to the strange quark case, and is due to
the relatively large strange quark mass needed for the overall fit to
the data. In addition, the R model calculations for dibaryon masses use
variational upper bounds for the kinetic energies, and so should lead
to overestimates in all cases. Finally, the two versions of the R model
differ from each other more with increasing strangeness since they
treat the flavor symmetry breaking in the one-body matrix elements
differently, as described earlier.

Excluding a few special cases, the mass differences, among all of these
model estimates are in the neighborhood of $\pm$50~MeV, on average.  We
take this as the uncertainty of the present dibaryon mass estimate,
using the R and N models altogether.

Although our primary interest has been in the $\alpha$=(003), the R
model predicts an H particle state very similar to that of Jaffe\cite{1},
which has already been extensively searched for. Experimental
results\cite{12} on doubly strange hypernuclei cast doubt on its existence
as a bound state, (although the issue of the relative binding energy of 
an H in a nucleus vs that of two $\Lambda$'s has not, to our knowledge,
yet been addressed seriously).  In the R model, it is $\sim$60~MeV lower
than the $\Lambda \Lambda$ threshold, and its wave function is almost a
pure flavor singlet as is Jaffe's.  However, since the N model finds
the H to be unbound by 35~MeV, this state is clearly sensitive to
dynamical details. Therefore, in our models, it cannot definitively
theoretically be concluded that the H exists as a bound state.

A preliminary relativistic calculation of the $\alpha$=(003) state was
reported in Ref.\cite{10}, where this state is denoted $d^{\star}$.
This $\Delta \Delta$ state has the largest binding energy in the $u, d$
and $s$ three flavor world. Because it is a nonstrange state, the mass
estimated by our model (both relativistic and nonrelativistic) should
be more reliable.  Even if we take the highest estimated mass, it is
still 15~MeV lower than NN$\pi\pi$ threshold. Therefore the $d^{\star}$
is quite possibly a narrow dibaryon resonance. For comparison, a
small-hard-core radius, large NN$\rho$-coupling meson exchange model
obtained a similar state\cite{13}. 

The strange sector also has a few dibaryon candidates.  We note
especially the spin~3 states, which all have a larger predicted binding
energy than the d$^*$.  They have been discussed in Ref.\cite{4}; we 
refer the reader to that discussion.

As to the question of prospects for experimental searches: There have,
of course, been many efforts to search for Jaffe's H. For the
$d^{\star}$ state, a preliminary estimate of the $\pi^{\pm}d
\rightarrow \pi^{\pm} d^{\star}$ production cross section is of order
0.1 $\mu$b\cite{10}.  This may be too small to be detected within the large
general pion scattering background. A similar reaction\cite{14}
\begin{equation}
\pi^{-} + ^{3}He \rightarrow n + d^{\star} \rightarrow n + n + n + \pi^{+}
\end{equation}
and its charge conjugate reaction
\begin{equation}
\pi^{+} + ^3H \rightarrow p + d^* \rightarrow p + p + p + \pi^-.
\end{equation}
may be more favorable for the detection of $d^{\star}$ production. Here
one can suppress the background by measuring the `spectator' $n (p)$
and the emitted $\pi^{+}(\pi^{-})$ in coincidence.  Unfortunately, 
pion beams of the required energy range and intensity do not seem to be
currently available at any accelerator facility.

On the one hand, coupling to the $I=0, \, ^{3}D_{3} \, NN $ channel is
expected to be small, obviating a search by this method. However on the
other hand, an analysis by Lomon\cite{earle} suggests that the
channel coupling may not be small enough to avoid conflict with results
from phase shift analyses of scattering data already available, even
though the experimental points are widely spread in energy.

Finally, we note that a recent calculation by Wong\cite{cw} of a
proton induced reaction similar to the pion induced one
\begin{equation}
p + ^{2}H \rightarrow p + d^{\star} \rightarrow p + p + p + \pi^{-}
\end{equation}
may produce sufficient signal for observation, again using detection of
the $\pi^{-}$ to suppress backgrounds. Such an experiment has been
proposed at TRIUMF (Canada) by S. Yen (Spokesman for E772).

{\noindent {\bf Acknowledgments} }

We wish to thank the members of the ``Further Directions in Pion
Physics'' study group at LAMPF, led by H.\ A.\ Thiessen, and including
G.\ Glass, C.\ Morris and E.\ Lomon, for stimulating discussions on the
favorable reactions for the production of the $d^{\star}$.

This work is supported by NSFC, the fundamental research fund of SSTC
and the graduate study fund of SEDC, both of China and by the
U.\ S.\ DOE.

\newpage

\begin{center}
Table 2a. Model A results: \\
An SU(3) flavor symmetric WF is used and the one body \\
energy difference due to $m_{s}\neq m$ has been neglected. \\
$s_0$ is the value of $x$ (in $fm.$) at which the minimum \\
energy of the dibaryon state occurs. \\

\vspace*{0.2in}

\begin{tabular}{|c|c|cllcc||l|} \hline
$SIJ$ &  & M$_{\alpha}$ & $V_{\alpha}$ & $\epsilon$ & $s_0$ &  
Threshold
   \\ \hline
001 & scs & 1902 & -8 & 0.2 & 1.4 & 1878(NN) \\
    & ccs & 1898 & -8 & 0.2 & 1.5 & \\

010 & scs & --- &   0 & 0.0 & 3.0 & 1878(NN) \\
    & ccs & --- &   0 & 0.0 & 3.0 & \\

003 & scs & 2143 & -349 & 1.0 & 1.3 & 2464($\Delta\Delta$) \\
    &   &  & &   &  & 2158(NN$\pi\pi$)  \\ \hline

-1$\frac{1}{2}$3 & scb & 2309 & -334 & 1.0 & 1.3 &  
2617($\Delta\Sigma^*$) \\
     & ccb & 2309 & -334 & 1.0 & 1.3 & 2335(N$\Lambda\pi\pi$) \\

-1$\frac{3}{2}$0 & scb & 2145 & -9 & 0.2 & 1.6 & 2132(N$\Sigma$) \\
    & ccb & 2145 & -9 & 0.2 & 1.6 &  \\
  \hline

-200 & scb & 2110 & -230 & 0.4 & 0.8 & 2231($\Lambda\Lambda$) \\
    & ccb & 2173 & -242 & 0.3 & 0.7 &  \\

-202 & scb & 2279 & -235 & 0.6 & 1.1 & 2472(N$\Xi^*$) \\
    & ccb & 2371 & -240 & 0.5 & 1.0 & 2397(N$\Xi\pi$) \\

-213 & scb & 2469 & -322 & 1.0 & 1.3 & 2765($\Delta\Xi^*$) \\
    & ccb & 2476 & -317 & 1.0 & 1.3 & 2690($\Delta\Xi\pi$) \\

-220 & scb & 2396 & -12 & 0.2 & 1.5 & 2386($\Sigma\Sigma$) \\
    & ccb & 2396 & -12 & 0.2 & 1.5 &  \\ \hline

-3$\frac{3}{2}$3 & scb & 2616 & -312 & 1.0 & 1.3 &  
2904($\Delta\Omega$) \\
    & ccb & 2634 & -302 & 1.0 & 1.3 & 2788($\Lambda\Xi^*\pi$) \\

-3$\frac{3}{2}$1 & scb & 2494 & -49 & 0.4 & 1.2 & 2511($\Sigma\Xi$) \\
  & ccb & 2496 & -50 & 0.4 & 1.2 &  \\

-3$\frac{1}{2}$2 & scb & 2405 & -255 & 0.6 & 1.0 & 2611(N$\Omega$) \\
  & ccb & 2481 & -252 & 0.4 & 0.9 & 2574($\Lambda\Xi\pi$) \\

-3$\frac{1}{2}$1 & scb & 2350 & -132 & 0.6 & 1.0 & 2434($\Lambda\Xi$) \\
  & ccb & 2456 & -139 & 0.4 & 0.9 &  \\
  \hline

-400 & scb & 2622 & -33 & 0.3 & 1.5 & 2636($\Xi\Xi$) \\
  & ccb & 2623 & -33 & 0.3 & 1.5 &  \\ \hline

-600 & scb & 3232 & -131 & 0.5 & 1.4 & 3345($\Omega\Omega$) \\
   \hline
\end{tabular}

\vspace*{0.2in}

$scs$ -- single channel only, flavor symmetry\\
$ccs$ -- with channel coupling, flavor symmetry\\
$scb$ -- single channel only, broken flavor symmetry\\
$ccb$ -- with channel coupling, broken flavor symmetry\\

\end{center}

\newpage

\begin{center}
Table 2b. Model B results: \\
One-body energy and the two-body matrix elements are both \\
calculated with the WF solutions for $m$=0 and $m_{s} = 307$~MeV. \\

\vspace*{0.2in}

\begin{tabular}{|c|c|cllcc||l|} \hline
$SIJ$ &  & M$_{\alpha}$ & $V_{\alpha}$ & $\epsilon$ & $s_0$ &  
Threshold
   \\ \hline
001 & scs & 1902 & -8 & 0.2 & 1.4 & 1878(NN) \\
    & ccs & 1898 & -8 & 0.2 & 1.5 & \\

010 & scs & --- &   0 & 0.0 & 3.0 & 1878(NN) \\
    & ccs & --- &   0 & 0.0 & 3.0 & \\

003 & scs & 2143 & -349 & 1.0 & 1.3 & 2464($\Delta\Delta$) \\
    &   &  & &   &  & 2158(NN$\pi\pi$)  \\ \hline

-1$\frac{1}{2}$3 & scb & 2293 & -354& 0.9 & 1.2 &  
2617($\Delta\Sigma^*$) \\
     & ccb & 2293 & -354 & 0.9 & 1.2 & 2335(N$\Lambda\pi\pi$) \\

-1$\frac{3}{2}$0 & scb & 2145 & -9 & 0.1 & 1.6 & 2132(N$\Sigma$) \\
    & ccb & 2145 & -9 & 0.1 & 1.6 &  \\
  \hline

-200 & scb & 2071 & -269 & 0.4 & 0.8 & 2231($\Lambda\Lambda$) \\
    & ccb & 2171 & -274 & 0.2 & 0.6 &  \\

-202 & scb & 2246 & -277 & 0.5 & 1.0 & 2472(N$\Xi^*$) \\
    & ccb & 2332 & -286 & 0.4 & 0.9 & 2397(N$\Xi\pi$) \\

-213 & scb & 2432 & -363 & 0.9 & 1.2 & 2765($\Delta\Xi^*$) \\
    & ccb & 2439 & -358 & 1.0 & 1.2 & 2690($\Delta\Xi\pi$) \\

-220 & scb & 2393 & -22 & 0.3 & 1.3 & 2386($\Sigma\Sigma$) \\
    & ccb & 2393 & -22 & 0.3 & 1.3 &  \\ \hline

-3$\frac{3}{2}$3 & scb & 2564 & -374 & 0.8 & 1.1 &  
2904($\Delta\Omega$) \\
    & ccb & 2581 & -363 & 0.7 & 1.1 & 2788($\Lambda\Xi^*\pi$) \\

-3$\frac{3}{2}$1 & scb & 2459 & -98 & 0.3 & 1.0 & 2511($\Sigma\Xi$) \\
  & ccb & 2462 & -99 & 0.3 & 1.0 &  \\

-3$\frac{1}{2}$2 & scb & 2346 & -326 & 0.6 & 0.9 & 2611(N$\Omega$) \\
  & ccb & 2420 & -327 & 0.4 & 0.8 & 2574($\Lambda\Xi\pi$) \\

-3$\frac{1}{2}$1 & scb & 2296 & -197 & 0.6 & 0.9 & 2434($\Lambda\Xi$) \\
  & ccb & 2420 & -210 & 0.3 & 0.8 &  \\
  \hline

-400 & scb & 2591 & -71 & 0.3 & 1.3 & 2636($\Xi\Xi$) \\
  & ccb & 2591 & -71 & 0.3 & 1.3 &  \\ \hline

-600 & scb & 3093 & -281 & 0.5 & 1.1 & 3345($\Omega\Omega$) \\
   \hline
\end{tabular}

\vspace*{0.2in}

Second column labels are same as in Table 2A. 

\end{center}

\newpage
\begin{center}
Table 3. The Comparison of the masses of six-quark system
between different versions of model.\\

\vspace*{0.2in}

\begin{tabular}{|c|ccc|ccc|ccc|ccc|} \hline
$SIJ$ & \multicolumn{3}{c|}{NA: $\nu = 1.6 fm^{-2}$} & 
        \multicolumn{3}{c|}{NB: $\nu = 0.6 fm^{-2}$}
      & \multicolumn{3}{c|}{RA} & \multicolumn{3}{c|}{RB} \\ \hline
 & M$_{\alpha}$ & $\epsilon$ & $s_0$ & 
   M$_{\alpha}$ & $\epsilon$ & $s_0$ & 
   M$_{\alpha}$ & $\epsilon$ & $s_0$ & 
   M$_{\alpha}$ & $\epsilon$ & $s_0$  \\ \hline
001 & 1874 & 0.2 & 1.3 & 1879 & 0.3 & 1.3 &
      1898 & 0.2 & 1.5 & 1898 & 0.2 & 1.5  \\

010 & 1885 & 0.1 & 1.6 & 1890 & 0.2 & 1.4 & 
      --- & 0.0 & 3.0 & --- & 0.0 & 3.0 \\

003 & 2084 & 1.0 & 1.2 & 2073 & 1.0 & 1.2 &
      2143 & 1.0 & 1.3 & 2143 & 1.0 & 1.3 \\ \hline

-1$\frac{1}{2}$3 & 2292 & 1.0 & 1.1 & 2285 & 1.0 & 1.1 &
      2309 & 1.0 & 1.3 & 2293 & 0.9 & 1.2  \\ 

-1$\frac{3}{2}$0 & 2131 & 0.2 & 1.4 & 2127 & 0.2 & 1.4 & 
      2145 & 0.2 & 1.6 & 2145 & 0.1 & 1.6  \\ \hline    

-200 & 2255 & 1.0 & 0.6 & 2257 & 1.0 & 0.6 & 
      2173 & 0.3 & 0.7 & 2171 & 0.2 & 0.6   \\

-202 & 2456 & 1.0 & 0.8 & 2455 & 1.0 & 0.8 &
      2371 & 0.5 & 1.0 & 2332 & 0.4 & 0.9 \\

-213 & 2508 & 1.0 & 1.0 & 2495 & 1.0 & 1.1 &
      2476 & 1.0 & 1.3 & 2439 & 1.0 & 1.2 \\

-220 & 2386 & 0.2 & 1.3 & 2382 & 0.2 & 1.3 & 
      2396 & 0.2 & 1.5 & 2393 & 0.3 & 1.3  \\ \hline

-3$\frac{3}{2}$3 & 2711 & 1.0 & 0.9 & 2697 & 1.0 & 1.0 &
      2634 & 1.0 & 1.3 & 2581 & 0.7 & 1.1 \\

-3$\frac{3}{2}$1 & 2509 & 0.4 & 1.0 & 2506 & 0.4 & 1.0 &
      2496 & 0.4 & 1.2 & 2462 & 0.3 & 1.0  \\

-3$\frac{1}{2}$2 & 2669 & 1.0 & 0.6 & 2670 & 1.0 & 0.6 &
      2481 & 0.4 & 0.9 & 2420 & 0.4 & 0.8  \\

-3$\frac{1}{2}$1 & 2560 & 1.0 & 0.6 & 2562 & 1.0 & 0.6 & 
      2456 & 0.4 & 0.9 & 2420 & 0.3 & 0.8  \\ \hline    

-400 & 2632 & 0.2 & 1.1 & 2630 & 0.3 & 1.1 &
      2623 & 0.3 & 1.5 & 2591 & 0.3 & 1.3  \\ \hline

-600 & 3375 & 1.0 & 0.6 & 3376 & 1.0 & 0.6 & 
      3232 & 0.5 & 1.4 & 3093 & 0.5 & 1.1 \\ \hline
\end{tabular}

\end{center}

\end{document}